# X-rays from Proton Bremsstrahlung: Evidence from Fusion Reactors and Its Implication in Astrophysics

Nie Luo
Department of Nuclear, Plasma and Radiological Engineering
University of Illinois, at Urbana-Champaign
Urbana, IL 61801, USA
nluo@illinois.edu

**ABSTRACT**

**In a fusion reactor, a proton and a neutron generated in previous reactions may again fuse with each other. Or they can in turn fuse with or be captured by an un-reacted deuteron. The average center-of-mass (COM) energy for such reaction is around 10 keV in a typical fusion reactor, but could be as low as 1 keV. At this low COM energy, the reacting nucleons are in an *s*-wave state in terms of their relative angular momentum. The single-gamma radiation process is thus strongly suppressed due to conservation laws. Instead the gamma ray released is likely to be accompanied by x-ray photons from a nuclear bremsstrahlung process. The x-ray thus generated has a continuous spectrum and peaks around a few hundred eV to a few keV. Therefore, the majority of this nuclear bremsstrahlung radiation is in the form of soft x-ray photons. The average photon energy and spectrum properties of such a process are calculated with a semiclassical approach. The results give a peak near 1.1 keV for the proton-deuteron fusion and a power-to-the-minus-second law in the spectrum's high-energy limit. The high-energy power law from nuclear bremsstrahlung is harder than that of the ordinary bremsstrahlung from electrons of a Maxwell distribution. The hard x-ray portion of this radiation is therefore not negligible compared to the thermal electron-bremsstrahlung type. An analysis of some prior tokamak discharge data shows that this phenomenon might have been observed before, and its interpretation is complicated by the presence of non-thermal electron bremsstrahlung. Nuclear bremsstrahlung in general and the proton type in particular may lead to new plasma diagnostics which are more sensitive to the ionic or nuclear degree of freedom. Such a prospect is helped by significant radiation in the form of hard x-rays of a few hundred keV. This phenomenon should also play a role in nuclear astrophysics as one of the sources for astrophysical x-rays. The process contributes particularly to stellar evolution in the early stage, where the temperature of proto-stars or the so called pre-main sequence stars (T Tauri stars, for example), is at a relatively low of several million degrees Kelvin. An order-of-magnitude calculation was made on the proton-deuteron fusion rate in young star objects. The estimated x-ray luminosity from this reaction is found enough in magnitude to account for experimental ones.**

# INTRODUCTION

For future thermonuclear reactors based on the D-T reaction, the generation of neutrons is the natural consequence of the major power producing reaction. However, protons are also generated by the parasitic D-D reaction according to

$$d + d \rightarrow {}^3\text{H} + p, \tag{1}$$

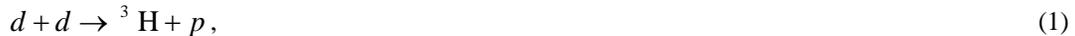

where *d* is a deuteron. This reaction has a small but non-negligible rate compared to the D-T reaction. In fact its rate is from one to two orders of magnitude within that of D-T between 10 keV and 100 keV.

The proton, once generated, could again participate in the thermonuclear process via other reactions. The simplest is the proton capture on neutron,

$$p + n \rightarrow {}^2\text{H} + \gamma. \tag{2}$$

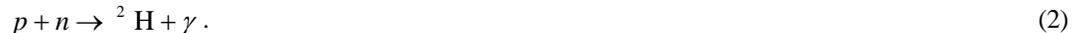

Alternatively, it can fuse with the un-reacted deuterium fuel via

$$p + d \rightarrow {}^3\text{He} + \gamma. \tag{3}$$

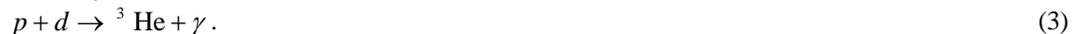

Processes in Eqs. (2) and (3) are well known to generate a gamma photon (2.23 MeV and 5.49 MeV respectively), with possible complications to the reactor vessel design. We now consider an associated or parasitic radiation accompanying Eqs. (2) and (3) due to a nuclear bremsstrahlung process, which will generate soft x-rays (one or multiple photons) on the order of one keV besides the MeV gamma. For example, the reaction (3) is more precisely

$$p + d \rightarrow {}^3\text{He} + \gamma + nX , \qquad (4)$$

where the total energy carried away by the gamma and the soft x-ray is 5.49 MeV, minus the small recoil kinetic energy of the $^3$He, which is often negligible. The number of photons is given in $n$. $n = 1$ should be more likely than the case of multiple photons. However, a clear-cut counting of photon number is difficult in a bremsstrahlung because the radiation is not of a single energy. Similarly reaction (2) is more appropriately

$$p + n \rightarrow {}^2\text{H} + \gamma + nX . \qquad (5)$$

The bremsstrahlung x-ray spectrum is continuous, and the photon released in this process is very soft, on the order of a few keV or less. Because soft x-ray of this energy is difficult to detect, and gamma ray in Eq. (4) is still very close to 5.49 MeV, this reaction is difficult to differentiate from the standard one in Eq. (3). In essence, the soft x-ray generation is due to the following physical process.

First take the simplest case of *p-n* capture as an illustration. A proton and a neutron attract each other via the nuclear (strong) force. The strong nuclear attraction causes both nuclei to accelerate to each other. Because the proton is charged, electromagnetic EM radiation is generated according to the theory of classical electrodynamics. The radiation is therefore similar to a bremsstrahlung in the nuclear domain. The release of the energy of the nuclei, is not through the deceleration alone, it also proceeds during *acceleration* by the strong force because the nucleons undergo both acceleration and deceleration in such a process. Here we loosely term it nuclear Bremsstrahlung, or maybe nuclear Startstrahlung to accentuate its origin in acceleration due to the nuclear force. Such a mechanism radiates x-ray photons around one keV, as we will demonstrate later. The case for proton-deuteron nuclear bremsstrahlung is in principle similar to that of the *p-n* type. An added complication is due to the Coulomb repulsion. However, once the Coulomb barrier is overcome by quantum mechanical tunneling, the strong nuclear attraction still causes both nuclei to accelerate to each other, albeit at different rates due to the mass difference. Both particles are charged and hence they all radiate EM quanta. Because both particles are positively charged but move in opposite directions, the radiation from them tends to cancel at the far field due to opposite accelerations. The cancellation is however not complete because the acceleration is not identical for proton and deuteron due to their mass difference.

The spectrum of such a radiation can be calculated in a semiclassical approach with the inclusion of quantum mechanical effects, as what we will demonstrate in the following section.

## SEMICLASSICAL RADIATION SPECTRUM

The nuclear bremsstrahlung radiation in *n-p* and *p-d* capture can be treated semiclassically due to the fact that the energy carried away by the soft photon, on the order of a few hundred eV to a few keV, is much smaller compared to the nuclear potential energy involved and the kinetic energy of the nucleons. This is equivalent to having a Planck constant $\hbar$ approaching 0 so that the quantum mechanical problem returns to a classical one. At the energy and dimensional scales of the problem, the wave nature of the nucleon dominates, and therefore the particle needs to be described by a wavepacket of appreciable spatial extension. However, the radiation sector is adequately handled by a semiclassical treatment of the nucleon acceleration in the nuclear field. The wavepacket nature of the nucleon motion smears out the sharp variation of the potential by convoluting the potential $V(r) = V_0/(e^{\mu r} - 1)$ with the wavefunction $\psi$ of the nucleon, which can be approximated by exponential functions in the coordinate space. The exact functional form of such a convolution is not very important, as we will see, and this type of acceleration in general gives rise to a $\omega^{-2}$ power law in the x-ray spectrum. As long as it is a short range attractive force, the answer is essentially the same. After all, this phenomenon is not a surprise because nuclear scattering at low energy is well known to be potential-shape independent.

A detailed semiclassical treatment of the problem in light of quantum mechnical conservation laws is discussed in detail in [1]. Here we briefly quote the similar results derived there, with a straightfoward adoption to the two cases of proton-neutron and proton-deuteron bremsstralung. The high frequency power spectrum is therefore given by

$$P(\omega) = -\frac{e^2}{6\pi^2\varepsilon_0 c^3}\frac{4kk_1^3}{(a-1)^{2.5}}\omega^{-2}. \tag{6}$$

The definition of coefficients is given in [1]. The spectrum at the long wavelength limit is dominated by a $\omega^3$ scaling law for the soft x-ray spectrum when the photon energy approaches 0. The derivation is straightforward and the result is given below

$$P(\omega) = \frac{e^2}{6\pi^2\varepsilon_0 c^3}\frac{\omega^3}{\mu^2}. \tag{7}$$

$\mu$ in Eq. (7) is given by a specific nuclear-nuclear potential and differs in the two cases of proton-neutron and proton-deuteron radiations. The $\omega^3$ long wavelength limit should intersect the high frequency behavior of Eq. (6) at a frequency $\omega$ given by

$$-\frac{e^2}{6\pi^2\varepsilon_0 c^3}\frac{4kk_1^3}{(a-1)^{2.5}}\omega^{-2} = \frac{e^2}{6\pi^2\varepsilon_0 c^3}\frac{\omega^3}{\mu^2}. \tag{8}$$

Such an $\omega$ approximately corresponds to the peak in the bremsstrahlung spectrum $\omega_p$. Solving the above equation for $\omega$, we obtain an approximate expression for the peak photon frequency

$$\omega_p \approx \omega = (\frac{1}{2})^{1/5}\frac{(2-a)^{\frac{3}{5}}}{(a-1)^{1.1}}\mu k^{0.5}. \tag{9}$$

When the total energy (kinetic + potential) of the proton is slightly lower than 0, it oscillates back and forth around the COM. The energy release into x-ray radiation is intermittent, being at maximum near the center and is negligible when far from there. The energy release in each cycle is given as,

$$\begin{aligned} E &= \frac{2e^2}{45\pi\varepsilon_0 c^3}k^{1.5}\mu \\ &\cong 34.004\,\text{keV}. \end{aligned} \tag{10}$$

Therefore, the energy released in each oscillation pass is roughly 34 keV. However notice that, the energy release here refers to a fictitious particle whose reduced mass is $m_p/2$, radiating around a point fixed in the inertial space. The relative motion of proton and neutron, is described by the preceding derivation but the acceleration of the proton in the inertial frame of references is only 1/2 of the calculation above. This results into an effective energy released of only 1/4. The energy of the photon is therefore only 8.5 keV. If it is interpreted as a release to two photons in order to conserve parity, the average photon energy will be very close to 4.25 keV.

It is reasonable to expect that the average photon energy of the preceding paragraph be close to the peak photon energy (frequency) derived in Eq. (9), or

$$\hbar(\frac{1}{2})^{1/5}\frac{(2-a)^{3/5}}{(a-1)^{1.1}}\mu k^{0.5} \approx 4.25\,\text{keV}. \tag{11}$$

This guides us to a characteristic spectrum of the *p-n* bremsstrahlung, as depicted in Fig. 1.

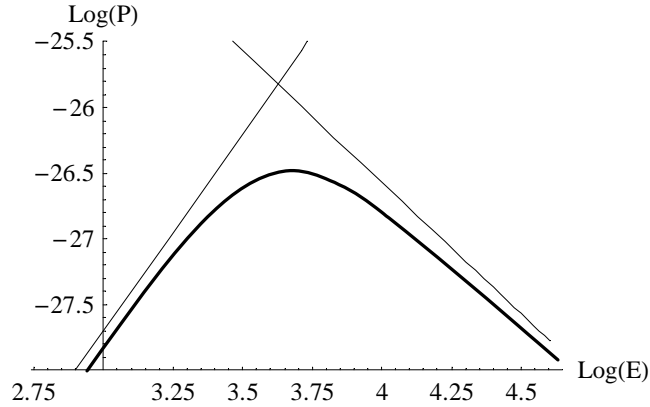

Figure 1. The bremsstrahlung x-rays spectrum from the radiative proton capture of neutron. The two straight lines are asymptotic for the low and high energy limits. The thick curve is the actual spectrum for intermediate energy. The horizontal axis is $\log_{10}(E)$ with $E$ the photon energy in eV. The vertical axis is $\log_{10}(P)$ with $P$ the spectral radiation power density in W per unit frequency. The p-n bremsstrahlung has a peak near the photon energy of 4.25 keV.

Although Fig. 1 is specific to the p-n nuclear interaction, the general shape and characteristics of the spectrum should apply to bremsstrahlung of other light nuclei at the keV-level energy. There could be more than one order of magnitude variation in the spectral radiation power density due to mass difference and the fact the radiation is inversely proportional to the mass squared. The average photon energy also decreases with increasing mass and this is why the process is mainly important to light nuclei.

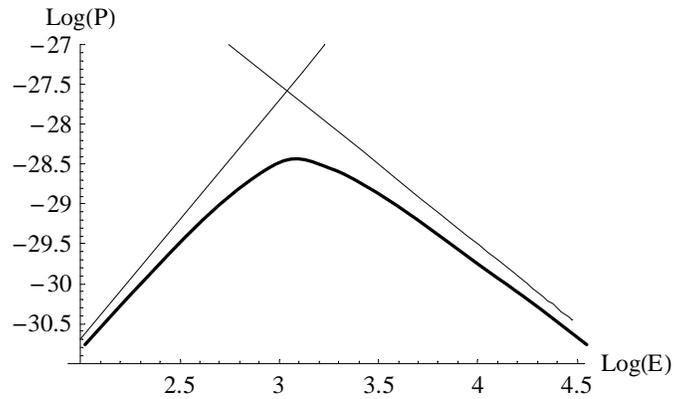

Figure 2. The bremsstrahlung x-rays spectrum due to the p-d fusion. The two straight lines are asymptotic for the low and high energy limits. The meaning of axes and labels are similar to those of Fig. 1. The p-d bremsstrahlung radiation has a peak near 1.1 keV. Later in this chapter the peak is often loosely referred to as 1-keV x-ray.

The spectrum for the p-d bremsstrahlung is similarly given and summarized in Fig. 2. It does not differ much from the p-n type shape-wise. In the following section we will see p-d nuclear bremsstrahlung is likely already observed in some prior tokamak discharge experiments, although it is often mixed with the so-called non-thermal electron bremsstrahlung. The peak bremsstrahlung of p-n is nearly two orders of magnitude higher than that of p-d, partly helped by the smaller reduced mass. Later in this chapter, we may refer soft radiation of this p-d bremsstrahlung type loosely as 1-keV x-ray, neglecting its minor difference from the 1.1-keV peak.

The experiments to be analyzed in the following section were not for neutron yield (i.e., tritium was not used) and therefore the *p-n* type soft x-ray is negligible. The experiments include deuterium discharge, which in principle could generate neutrons too via the reaction $d+d \rightarrow {}^3\text{He}+n$. However, this *neutron* generation is weighted down by the Coulomb penetration factor at low energy of a few keV. Fusion of *p-d* is weighted down too by a similarly sized Coulomb barrier, but the penetration (tunneling) factor is orders of magnitude higher at 1 keV, helped by the smaller reduced mass in the *p-d* system (0.67 versus 1 in *d-d,* in terms of the proton mass). Even if the neutron is generated, the probability of colliding with a proton is far below 1 for an ordinary tokamak. Overall, it is not difficult to prove that bremsstrahlung from *p-d* fusion is more important in the discharge experiments to be discussed.

# EXPERIMENTAL EVIDENCE

The soft x-ray spectrum due to a nuclear bremsstrahlung of proton-deuteron fusion assumes a power law in its high energy tail. This is entirely different from the exponential behavior expected from the bremsstrahlung spectrum of thermal electrons.

Ordinary bremsstrahlung from the electron-ion scattering should have an exponential drop-off tail when the photon energy is significantly higher than the plasma temperature, say after a few keV. One wants a lucid picture as to why nuclear bremsstrahlung instead has a power law. Simply put, the electron bremsstrahlung from an ion is due to a repulsive force with a characteristic energy scale not more than the initial kinetic energy of the electron. In comparison, the nuclear bremsstrahlung is essentially from an *attractive* strong force, whose energy scale is on the order of few tens of MeV. The energy available to radiate is at least that of nuclear binding energy (BE) of a few MeV or more even if the nucleons are initially at a zero kinetic energy. If both nuclei have initial kinetic energy (KE), the radiated photon can be as high as KE + BE, explaining the high energy tail of the power law spectrum.

Non-thermal x-ray behaviors have been reported before from fusion reactors, in particular, tokamaks [2,3]. It has been traditionally attributed to various atomic processes. However as we can see in this study, the interpretation of the tokamak non-thermal x-ray, which also has a power law at the high energy limit, is complicated by the presence of nuclear bremsstrahlung x-ray. Alternatively one might ask how much portion of the observed non-thermal x-ray flux is really due to atomic processes and other electron run-away phenomena, with the rest of radiation being due to nuclear bremsstrahlung. A detailed discussion of the phenomenon and its explanation are given in [4], but a quick epitome is summarized as follows for the sake of convenience.

In the interesting experiment reported in [3], there is a puzzling discrepancy in the strength of non-thermal x-rays from hydrogen and deuterium discharges. The non-thermal x-ray is phenomenologically modeled by a fitted electron distribution function. However, there is no compelling reason as to why deuterium and hydrogen should have such a different non-thermal x-ray yield if the radiation is only from electrons.

The electron bremsstrahlung spectrum is routinely given by

$$P(\nu) = 4.163 \times 10^{-26} n_e n_i (\frac{13.59\text{eV}}{kT})^{1/2} e^{-h\nu/kT} \overline{g}, \tag{12}$$

where $P(\nu)$ is in (erg/eV) cm$^{-3}$sec$^{-1}$. $\overline{g}$ is the Gaunt factor and is in general well approximated by:

$$\overline{g} = (\frac{T_e}{h\nu})^{1/3}.$$

Due to the power of one third, the Gaunt factor has relatively weak dependence on the electron temperature and photon energy, and is hence often taken as a constant. This is the approximation made in [3] in their data analysis. To be consistent with their analysis we follow this convention and hence the non-thermal x-ray flux at energy between 20 and 35 keV, i.e., in Fig. 7 of [3], is modeled by an electron-bremsstrahlung type of

$$P(\nu) = An_t (\frac{1}{T})^{1/2} e^{-h\nu/kT}, \tag{13}$$

where $A = 4.163 \times 10^{-26} n_i (\frac{13.59 \text{eV}}{k})^{1/2}$, and $n_t$ is the so-called effective "tail" electron density. Using Fig. 12 of [3], one gets the tail temperature $T_T$ of 7 and 21 keV for the hydrogen and deuterium discharge, respectively, at $\bar{n}_e = 1.6 \times 10^{14} \text{cm}^{-3}$. The photon energy is a bit arbitrary, but it is reasonable to use the average non-thermal of 27 keV shown in their Fig. 7. The experimental "tail" x-ray ratio between the hydrogen and deuterium discharge is therefore

$$\frac{P_D(\nu)}{P_H(\nu)} = (\frac{T_{TH}}{T_{TD}})^{1/2} \exp[\frac{-h\nu}{1}(\frac{1}{T_{TD}} - \frac{1}{T_{TH}})]$$
$$= (\frac{1}{3})^{1/2} \exp[\frac{-27}{1}(\frac{1}{21} - \frac{1}{7})] \qquad (14)$$
$$= 7.6$$

Overall the experimental non-thermal x-ray flux in deuterium discharge is roughly 7~8 times more than that of the hydrogen one. This can be at least partially explained by the nuclear bremsstrahlung process accompanying the *p-d* fusion process in the tokamak experiment. We know that natural hydrogen is in fact not 100% pure. Instead it contains one deuterium for every ~7000 hydrogen atoms. The tokamak hydrogen discharge therefore will actually burn deuterium with *p-d* fusion. The *p-d* fusion is then accompanied by significant non-thermal nuclear bremsstrahlung radiation when the nuclear kinetic energy is on the order of keV, following the reasoning that we have outlined so far. Similarly the deuterium discharge is not free from *p-d* fusion either due to impurity of hydrogen. A typical high-grade deuterium commercially available is 99.9% in D. In other words it has one hydrogen for every 1000 deuterium atoms. The *p-d* fusion rate is proportional to $n_D n_H$, or the product of deuterium and hydrogen density. Therefore a deuterium discharge is likely to have a non-thermal x-ray output 7 times that of a hydrogen one. The good match between the experiment and our proposed explanation indicates that the nuclear bremsstrahlung scenario is one likely reason for the observed anomaly.

One interesting aspect of the phenomenon is possibly new ways to monitor the fusion reaction as a new probe due to its intrinsic nuclear nature. The flux of the nuclear bremsstrahlung x-ray, for example, is more intimately associated with the ionic (nuclear) temperature rather than electron ones. This may open up new possibilities for plasma diagnostics in future fusion reactors.

# IMPLICATION FOR NUCLEAR ASTROPHYSICS

The nuclear bremsstrahlung outlined above represents an additional energy loss process accompanying nuclear fusion of light nuclei, and the net result is an increased fusion cross-section. Fusion cross-section is written as

$$\sigma(E) = \frac{S(E)}{E} e^{-2\pi\eta}, \qquad (15)$$

where $S(E)$ is the astrophysical factor related to the intrinsic nuclear process. $e^{-2\pi\eta}$ is the quantum mechanical tunneling rate of the Coulomb barrier, with $\eta$ given by

$$2\pi\eta = 31.29 Z_1 Z_2 (m_r/E)^{1/2}, \qquad (16)$$

where $Z_1$ and $Z_2$ are the atomic charge of relevant nuclei in the fusion process and $m_r$ the reduced mass in amu. $E$ is in keV in Eq. (16).

What Eq. (15) expresses is that the fusion cross-section would be inversely proportional to the COM energy $E$, if both the intrinsic nuclear cross-section and the Coulomb penetration factor are held constant. Therefore, losing more COM energy due to the nuclear bremsstrahlung process should result in a higher capture or fusion rate. Note that the nuclear bremsstrahlung only occurs significantly with the strong nuclear force, i.e., *after* the Coulomb barrier is penetrated, therefore it does not reduce the penetration probability in any notable way, because the energy loss during the tunneling is negligible. The astrophysical factor $S(E)$ is a slowly varying function of $E$, and the overall effect of the nuclear bremsstrahlung is a much increased fusion cross-section for nuclear fusions at an initial COM kinetic energy of a few hundred eV. Since each pass of the proton near the other nuclei, when attracted via nuclear force, more or less radiates about one keV, the effect is only significant when the initial COM energy is below 10 keV. Note that the amount of energy radiated is not affected appreciably by the COM energy until a very high COM energy (say over 50 MeV) is reached.

The effect is especially important to *p-d* fusion because the corresponding reduced mass is small, giving rise to strong radiation and hence large energy loss. The small reduced mass also makes the Coulomb penetration factor large, from Eq. (16).

For *p-d* fusion $m_r$ is $\approx 0.67$ compared to 1 for *d-d*. Then the *p-d* fusion is the favored reaction at a few hundred eV, even if the boost in fusion cross-section due to the energy loss from nuclear bremsstrahlung is not considered. All of the aforementioned factors combine to explain why the *p-d* plays a critical role in the evolution of proto-stars and other pre-main sequence stars, when the temperature is still at a low of several million Kelvin.

Recently, x-ray surveys of pre-main sequence stars confirmed elevated x-ray activities on many of the so-called T Tauri stars (TTSs) [5]. A T Tauri star is a fairly representative pre-main sequence star. This x-ray phenomenon from TTSs leaves a few questions not well answered so far. The first is concerned with the origin of the x-ray activity, e.g., what physics processes lead to the x-ray radiation of TTSs? And second, why do the TTSs have higher x-ray luminosities (up to ~$10^4$ times higher than those seen from the Sun)? Nuclear bremsstrahlung radiation from the *p-d* fusion process offers a quite logical explanation to these two questions.

Although the x-ray from TTSs can be easily attributed to shock and cold plasma due to disk accretion, as is sometimes quoted in the literature [6], the detailed physical process is in general not well understood. Moreover, many recent experiments have demonstrated the x-ray as the result of coronal activity, but not a direct outcome of accretion, at least for a certain category of TTSs or pre-main stars [5,7,8]. Therefore, the x-ray should come from some fairly general physics principles. The nuclear bremsstrahlung process in *p-d* fusion underscores a simple yet profound electrodynamics principle: accelerated charges radiate. Therefore, it constitutes an alternate yet fairly plausible explanation for the TTSs' x-ray flux. This does not come as a surprise after all if we consider the importance of *p-d* fusion (deuterium burning) in the evolution of proto-stars.

The proto-stars are well known to undergo the so-called deuterium burning process when the temperature in their centers has risen to the $10^6$ K scale due to initial gravity contraction in the proto-stellar accretion process [9,10]. Deuterium burning, or in fact the *p-d* fusion, acts as a central thermostat that keeps the proto-stellar temperature not much higher than a few million degrees Kelvin, and this should form an ideal nuclear bremsstrahlung x-ray radiation source according to the aforementioned analyses. When the x-ray is not obscured by dusts/clouds, which should happen in cases where deuterium burns at the stellar surface, i.e., in case of D-shell burning [11,12], or in the convective outer layer, strong x-ray radiation is detected and this should lead to the pre-main sequence stars being registered as a T Tauri type. This often happens at a relatively late stage in the TTSs evolution timeline. The deuterium consumed is of a primordial nature and the thermonuclear burning should use up much of D in the process, which lasts up to a few million years. Then these pre-main stars start to enter main-sequence with the primordial deuterium exhausted, radiating at a much lower x-ray luminosity.

The appearance of a TTS is generally preceded by a proto-star stage. Although the two are generally distinguished by their spectral energy distribution (SED), there is hardly a clear-cut borderline between the two stages. While deuterium burning is well accepted as the key in proto-stellar evolution, its role is less probed and mentioned in pre-main sequence stars, with the TTS included. However, there should not be surprise if deuterium burning still plays a role (though minor) in the pre-main sequence stars, concerning the accretion disks that roughly half of the TTSs have: the accretion process then keeps bringing in primordial deuterium and this results in *p-d* fusions near the TTS photosphere. Note that the *p-d* fusion likely does not occur directly in the photosphere due to its low temperature.

One may ask how to explain the enhanced x-ray from TTSs that are not accreting. In these cases, the x-ray is more of the coronal type like that in the Sun. The explanation could be fairly simple: the non-accreting TTSs are *in fact* non-obscured (or naked) proto-stars that still undergo deuterium burning in the D-shell or the convection zone. The existence of deuterium burning in pre-main sequence stars is strongly supported by recent x-ray observations [13]. If the accretion has stopped, the surface x-ray activity is no longer much masked by dusts. The deuterium burning process itself causes convection and differential rotation in outer stellar spheres. Strong magnetic field and starspots are generated as the result. The magnetic activity in turn will further reinforce deuterium-burning in the stellar corona due to the magnetic confinement effect. The *p-d* fusion rate is increased, not much different from similar nuclear reactions in terrestrial tokamaks. This process lasts until most of primordial deuterium is consumed.

Now we need a more quantitative study of deuterium-burning in proto-stars/TTSs and its effect on luminosity. Inserting the relevant parameters of *p-d* capture (i.e., the astrophysical factor, atomic charges, and the reduced mass) into Eqs. (15) and (16), we obtain its fusion cross-section as a function of temperature. However, density profile of a proto-star or a TTS is not precisely known, and therefore an accurate account of deuterium burning rate is not easily reached. Such details are not necessary for an order of magnitude (OOM) estimate though, which is the case here for this chapter: Our purpose here is mainly to see whether or not the *p-d* fusion power output can surpass the observed x-ray luminosity. The exponential (Gamow) factor of Eq. (15), after convoluting with the Maxwell distribution, can be approximated by a power law for a narrow energy window. This simplification leads to a fairly straightforward expression of power density for the *p-d* fusion [10],

$$W_D \approx 4\times 10^7 [D/H] \left(\frac{\rho}{\text{g}\cdot\text{cm}^{-3}}\right)\left(\frac{T}{10^6 \text{K}}\right)^{11.8} \text{erg}\cdot\text{g}^{-1}\cdot\text{s}^{-1} \quad . \tag{17}$$

We take the approximate primordial deuterium abundance of $2\times 10^{-5}$ for the D/H ratio. The stellar density is calculated with a typical TTS mass of $3M_\odot$ ($M_\odot$: solar mass) and radius of $6R_\odot$ ($R_\odot$: solar radius) with a simplified average [11]. This gives 0.02 g/cm$^3$ for $\rho$. The total luminosity due to *p-d* fusion is therefore calculated to be $9.36\times 10^{34}$ erg/s.

Clearly, the total luminosity from *p-d* fusion alone is more than one order of magnitude higher than that of the Sun, even at a very moderate temperature of $10^6$ Kelvin (1MK). If the T increases to 1.2MK ($1.2\times 10^6$ K), the total luminosity increases to $8\times 10^{35}$ erg/s. The luminosities at different temperatures are tabulated in Table I.

Now we need to find out what portion of the *p-d* fusion luminosity is in soft x-rays. From Eq. (4) the release of each 5.49-MeV gamma is accompanied by at least one 1-keV soft x-ray photon. In reality, multi-photon x-ray radiation could also happen. Therefore, the $n=1$ case, as defined in Eq. (4), serves as a conservative estimate on the x-ray radiation flux. Multiplying 1keV/5.49MeV with the total luminosities that we calculated before, the lower bound on x-ray luminosity is obtained and listed in the third column of Table I. An estimated x-ray luminosity of $1\times 10^{32}$ erg/s appears to be a good rule-of-thumb guidance.

Table I. Luminosities caused by *p-d* fusion at different temperatures for a TTS of 3M$_\odot$ and 6R$_\odot$.

| T (MK) | Total luminosity $L_{tot}$ ($10^{34}$ erg/s) | X-ray luminosity $L_{x1}$ (lower bound) ($10^{31}$ erg/s) | X-ray luminosity $L_{xr}$ (most probable) ($10^{32}$ erg/s) |
|---|---|---|---|
| 1.0 | 9.36 | 1.70 | 1.63 |
| 1.2 | 80.5 | 14.7 | 14.1 |
| 1.3 | 207 | 37.7 | 36.2 |

A more realistic and probable estimate of the average $n$ for the 1-keV soft x-ray is given as follows. Here we quote a well known result for bremsstrahlung from relativistic quantum mechanics [14]:

$$d\sigma(k) \propto \frac{dk}{k}, \tag{18}$$

where $k$ is the wave-vector of the bremsstrahlung photon.

A straightforward integration of Eq. (18) gives

$$\sigma(k) \propto \ln(k_{max}) - \ln(k), \tag{19}$$

where $k_{max}$ is the maximal wave-vector possible with the radiation. It usually has an upper limited required energy conservation.

Due to the linear relation between $k$ and $\omega$, the photon frequency, alternatively the above relation is written

$$\sigma(\omega) \propto \ln(5.49\text{MeV}) - \ln(\hbar\omega) + c, \tag{20}$$

where c is the integration constant, and can be set to 1, because we take the 5.49-MeV gamma radiation rate as unit. Also in Eq. (20) we have taken a reasonable upper limit of integration as the p-d binding energy.

The average number of 1-keV soft x-ray photon for each 5.49-MeV photon generated is then

$$n = \ln(\frac{5.49\text{MeV}}{\hbar\omega}) + 1 = \ln(\frac{5490}{1}) + 1 = 9.6. \tag{21}$$

This average $n = 9.6$ is then multiplied with the third column of Table I to give the more realistic x-ray luminosity, listed in the fourth column.

From Eq. (18), the experimental ratio between soft x-rays and the hard gamma also depends on the spectrometer resolution due to the factor $dk$. The number quoted in Eq. (21) therefore should be only taken as an OOM estimate.

One other assumption we made in the above estimate is that the energy dissipation in the *p-d* capture is only from electric dipole type acceleration, or bremsstrahlung. This is of course not the only radiative interaction. The other well-known energy loss is from magnetic dipole radiation [15]. Nevertheless, the conclusion that we draw from preceding analysis is clear: the addition of nuclear bremsstrahlung, so far neglected in most *p-d* rate calculation, could result in an x-ray flux much higher than that of gamma.

In practical experiments, the measured x-ray intensity is likely complicated by x-ray opacity: most of x-rays generated under the photosphere do not escape the TTS surface. The radiation could be further obscured by the accreting dusts/clouds. A precise account of such effects is difficult at this moment due to additional processes: For example, gravitation attraction experienced in the accretion process also help heat up the TTS atmosphere/surface, and ignite proton-deuterium fusion near but above the TTS' photosphere, where the resulting x-ray radiation should be observed relatively easily.

An additional source of x-rays from the TTS likely lies in its corona. The primordial deuterium in a TTS corona is less consumed than that in a main sequence star. Moreover, the disk accretion, if any, would bring in more primordial deuterium and help generate intense x-rays in the TTS corona.

Overall, the x-ray luminosity from *p-d* fusion in TTSs is significant. It could be over 5 orders of magnitude higher than that seen from the Sun, if not screened. Due to our limited knowledge in details (such as accretion heating, accretion obstruction, local structures like flares) about the distant TTSs, it is not entirely clear at this stage which portion of observed x-rays are from *p-d* fusion. But one thing is quite evident: there is so far no evidence strong enough to rule out the *p-d* fusion as a contributor of TTS x-ray radiation.

The characteristic spectrum of the TTS x-ray, which has a peak around 0.9 ~1.0 keV [16], is logically explained by the *p-d* fusion x-ray spectrum which peaks around 1.1 keV shown in Fig. 2. The slight softening or red-shift of the peak can be fairly logically accounted for by dust/cloud absorption. On the other hand, it is difficult to explain the x-ray peak from the traditional cold plasma radiation, whose temperature is not over 1MK. The radiation from a standard 1MK plasma should peak around at an energy much low than ~100 eV if it is not at zero.

The plasma radiation spectrum due to electron bremsstrahlung given in Eq. (12) is simplified to

$$P(\nu) \propto e^{-h\nu/kT} \left(\frac{kT}{h\nu}\right)^{1/3}, \tag{22}$$

where $T$ is the plasma temperature. Differentiating the right hand side of Eq. (22) with respect to $h\nu$ and making it zero, one reaches the energy corresponding to one extremum of $P(\nu)$

$$h\nu = -kT/3.$$

Its value is negative and therefore there is not any peak for any real (positive) temperature. Alternatively one can see that Eq. (18) is a monotonically decaying function of positive $h\nu$, and should only peak at a zero photon energy (frequency). This is clearly different from the experimental spectrum from TTSs.

Similarly x-ray spectrum with that so-called non-thermal (hot) tail is found from terrestrial fusion reactors and has been briefly explained in terms of *p-d* nuclear bremsstrahlung by the author in the previous section: EXPERIMENTAL EVIDENCE. The qualitative difference between a traditional plasma (electron bremsstrahlung) and the nuclear bremsstrahlung is already mentioned before and is once again emphasized as follows. Ordinary bremsstrahlung from electron-ion scatterings should have an exponential drop-off tail when the photon energy is significantly higher than the plasma temperature which is a few hundred eV for many stellar astrophysical processes. The drop-off is essentially given by the exponential factor from the Maxwell distribution of the electron kinetic energy because the x-ray radiation is from the conversion of electron kinetic energy. On the other hand, nuclear bremsstrahlung is essentially from an *attractive* strong force, whose energy scale is on the order of few tens of MeV. The energy radiated is not from the initial kinetic energy of the proton, but rather from the *p-d* nuclear potential energy. Therefore, the Maxwell exponential factor does not appear in the power spectrum.

Not surprisingly, many TTS x-ray spectra are fit by multi-temperature plasma models, very similar to the case of tokamak x-rays. The feature common to both astrophysics and terrestrial fusion x-ray studies seems to be the need of high energy component to make up the deficiency of a simple plasma radiation model that considering only the electronic and ionic degrees of freedom. As from what we have learned so far, this high temperature portion may not come from non-thermal high energy electrons after all. It could instead be the manifestation of strong nuclear force, whose energy scale is of course fairly high.

Therefore, the nuclear bremsstrahlung mechanism gives plausible answers to both of the TTSs questions, at least at an order of magnitude level. Whether or not it can explain the T Tauri star x-ray properties more quantitatively remain to be seen. More works are underway and results will be published elsewhere. One thing is clear: the experimental data so far cannot rule out nuclear bremsstrahlung from *p-d* fusion as one source of the observed x-rays.

Because fusion cross-sections are quite small for the terrestrial based fusion experiments at a few hundred eV where the nuclear bremsstrahlung effect is expected to be most important, the corroboration between the fusion reactor data and nuclear astrophysical ones is a synergistic way of verifying theories and guiding experiments. The large scale of astrophysical phenomena in a sense much compensate for the painfully small nuclear fusion cross-sections. Therefore, the data from nuclear astrophysics can often serve as a reference for studying aneutronic fusion cross-sections in terrestrial reactors, especially at low energy. The refinement on fusion cross-sections itself, obtained from nuclear experiments on the Earth, can in turn be fed back positively into astrophysical studies so as to better understand a variety of stellar processes.

# CONCLUSION

A semiclassical theory was developed to account for nuclear bremsstrahlung of light nuclei in the process of fusion and capture reactions. The focus was on proton bremsstrahlung on deuterons because this gives the strongest radiation among fusion reactions due to the lightness of proton. Some existing tokamak data were analyzed, indicating the important role of nuclear bremsstrahlung to fully understand some phenomena not well explained before. It is also a potential candidate for fusion plasma diagnostics at the photon energy of a few hundred keV. The theory was also adopted to explain the abnormally strong x-ray radiation from some pre-main sequence stars, and was shown to give qualitatively satisfactory accounts.